\documentstyle[12pt]{article}
\newcommand{\beq}{\begin{equation}}
\newcommand{\eeq}{\end{equation}}
\newcommand{\bea}{\begin{eqnarray}}
\newcommand{\eea}{\end{eqnarray}}

\renewcommand{\b}{\beta}

\newcommand{\m}{\mu}

\newcommand{\s}{\sigma}

\newcommand{\oh}{\frac{1}{2}}

\newcommand{\ot}{\frac{3}{2}}
\newcommand{\dg}{\dagger}
\newcommand{\non}{\nonumber}

\renewcommand{\t}{\tau}
\newcommand{\rf}[1]{(\ref{#1})}
\newcommand{\ra}{\rightarrow}

\begin{document}

\addtolength{\baselineskip}{0.20\baselineskip}

\hfill hep-lat/9510028

\hfill SWAT/95/91

\hfill LBL-37855

\hfill October 1995

\begin{center}

\vspace{48pt}

{ {\bf Casimir Scaling vs. Abelian Dominance
             in QCD String Formation} }

\end{center}

\vspace{18pt}

\begin{center}
{\sl L. Del Debbio${}^a$, M. Faber${}^b$, J. Greensite${}^{c,d}$,
and {\v S}. Olejn\'{\i}k${}^e$}

\end{center}

\vspace{32pt}

\begin{tabbing}

{}~~~~~~~~~~~~~\= blah  \kill
\> ${}^a$Department of Physics, University of Wales Swansea, \\
\> ~SA2 8PP Swansea, UK.  E-mail: {\tt L.Del-Debbio@swansea.ac.uk} \\
\\
\> ${}^b$Institut f\"ur Kernphysik, Technische Universit\"at Wien, \\
\> ~1040 Vienna, Austria.  E-mail: {\tt faber@kph.tuwien.ac.at} \\
\\
\> ${}^c$Physics and Astronomy Dept., San Francisco State University, \\
\> ~San Francisco CA 94132 USA.  E-mail: {\tt greensit@stars.sfsu.edu} \\
\\
\> ${}^d$Theory Group, Lawrence Berkeley Laboratory, \\
\> ~Berkeley CA  94720 USA.  E-mail: {\tt greensite@theorm.lbl.gov} \\
\\
\> ${}^e$Institute of Physics, Slovak Academy of Sciences, \\
\> ~842 28 Bratislava, Slovakia.  E-mail: {\tt fyziolej@savba.savba.sk} \\

\end{tabbing}

\newpage

\noindent {~}

\vspace{48pt}

\begin{center}

{\bf Abstract}

\end{center}

\bigskip

    We show that the hypothesis of abelian dominance in maximal
abelian gauge, which is known to work for Wilson loops in the
fundamental representation, fails for Wilson loops in higher group
representations. Monte Carlo simulations are performed on lattice
$SU(2)$ gauge theory, in $D=3$ dimensions, in the maximal abelian gauge, in
the confined phase.  It is well-known that Creutz ratios extracted from
loops in various group representations are proportional to the quadratic
Casimir of each representation, in a distance interval from the confinement
scale to the point where color screening sets in.  In contrast we find
numerically, in the same interval, that string tensions extracted from loops
built from abelian projected configurations are the same for the fundamental
and $j=3/2$ representations, and vanish for the adjoint representation.  In
addition, we perform a lattice Monte Carlo simulation of the Georgi-Glashow
model in $D=3$ dimensions. We find that the representation-dependence of
string tensions is that of pure Yang-Mills in the symmetric phase, but changes
abruptly to equal tensions for the $j=1/2,~3/2$ representations, and zero
tension for $j=1$, at the transition to the Higgs phase.  Our results
indicate that an effective abelian theory at the confinement scale,
invoking {\it only} degrees of freedom (monopoles and photons)
associated with a particular Cartan subalgebra, is inadequate to describe
the actual interquark potential in an unbroken non-abelian gauge theory.

\vfill

\newpage

\section{Introduction}

   Many years ago, in a very influential paper \cite{Poly}, Polyakov
demonstrated quark confinement in the Higgs phase of the $D=3$
Georgi-Glashow model, the mechanism being condensation of monopoles
in the unbroken $U(1)$ subgroup.  It is natural to suppose that such
a mechanism might also explain quark confinement in the symmetric phase
of non-abelian gauge theories, in $D=4$ as well as $D=3$ dimensions, and
that the effective theory at the confinement scale and beyond is essentially
abelian, i.e. compact QED. The most explicit version of
this idea is the abelian projection theory due to 't Hooft \cite{thooft},
where a special gauge-fixing condition on the gauge fields, rather than
the Higgs field, is used to single out an abelian subgroup of the
full gauge group.  For an $SU(N)$ theory, 't Hooft's abelian projection
gauge-fixing leaves an unbroken $U(1)^{N-1}$ subgroup; condensation of
the magnetic monopoles associated with this subgroup is the conjectured
confinement mechanism.  This picture is one possible realization of
the idea of dual superconductivity in non-abelian gauge theories, as
originally proposed by 't Hooft \cite{th2} and Mandelstam \cite{Mand}.

    In the $D=3$ Georgi-Glashow model ($GG_3$) in the Higgs phase,
Polyakov computed the area law contribution to Wilson loops in terms of
an effective abelian theory, invoking only the monopoles and ``photons''
associated with the unbroken $U(1)$ gauge group.  The abelian gauge field
($A_\m^3$, say) is singled out by a unitary gauge choice, and for the
calculation of the string tension (in this theory) it is a reasonable
approximation to ignore the contribution of the other color components,
i.e.
\bea
       <W(C)> &=& <\mbox{Tr}\exp[i\oint dx^\m A^a_\m \t_a]>
\non \\
              &\sim& <\mbox{Tr}\exp[i\oint dx^\m A^3_\m \t_3]>
\label{abdom}
\eea
where $\t_a = \oh \s_a$.
The same approximation, in the context of 't Hooft's theory, has
come to be known as ``abelian dominance'' \cite{Suzuki}.

   In this article we address the question of whether abelian dominance,
which implies the existence of an effective abelian theory of monopoles
and photons at large scales, is
adequate to describe the infrared dynamics of $D=3$ Yang-Mills theory,
in the maximal abelian gauge. Our tool for studying this question
will be Wilson loops in higher group representations.  It should be
noted, at the outset, that we are {\it not} addressing the possible
relevance or irrelevance of monopoles, or the validity of
dual-superconductor pictures in general. Our investigation is limited
to one issue only, namely: are vacuum
fluctuations, at the confinement scale and beyond, dominated by fluctuations
in the gauge field associated with a Cartan subalgebra of the gauge
group, as is the case for $GG_3$ in the Higgs phase?  In
particular, the question here is not
whether magnetic monopoles, defined with respect to an abelian projection
gauge, are condensed (evidence for condensation of such monopoles is found
in ref. \cite{Luigi}; condensation of these and perhaps other types of
field configurations is not unexpected in a magnetic-disordered
vacuum state).  Rather, the issue we address is whether the fluctuations
of the corresponding (Cartan subalgebra) gauge
field dominate the large-scale vacuum fluctuations,
justifying the use of eq. \rf{abdom}.

  There are a number of reasons to believe in abelian dominance
for pure Yang-Mills theory in maximal abelian gauge. There are,
for example, several kinematical similarities between that
theory, and $GG_3$ in the Higgs phase.  First, in both cases,
the underlying $SU(2)$ symmetry is reduced to a $U(1)$ symmetry by
a gauge choice: the unitary gauge in $GG_3$, and the maximal-abelian gauge
\cite{KLSW} for pure Yang-Mills.  Second, magnetic monopoles can be
identified in both theories, associated with the remaining $U(1)$ symmetry.
Third, on the lattice, one finds in both cases that most of the quantum
fluctuations of the link variables are in the $A^3_\m$ degrees of freedom.
Apart from these kinematical similarities, it is reasonable to suppose that
if abelian monopoles are the crucial confining configurations, then
a truncation to the associated $A^3_\m$
degrees of freedom (abelian dominance) would retain the essential
features of magnetic disorder and flux-tube formation.  In support
of this supposition, Monte Carlo simulations have found that
the abelian dominance approximation, i.e. eq. \rf{abdom}, accurately
reproduces the string tension for Wilson loops in the fundamental
representation of the gauge group \cite{Suzuki}.

   However, the fundamental representation is not the only group
representation, and Wilson loops in higher
group representations may also have a tale to tell.   In particular
let us recall the suggestion, made many years ago, that the string tension of
planar Wilson loops in $D=3$ and $D=4$ dimensions could be computed from
an effective 2-dimensional gauge theory.  This suggestion, known as
``dimensional reduction'', was put forward independently (and for quite
different reasons) in ref. \cite{JG} and \cite{Poul}, and some numerical
evidence for the idea, based on a Monte Carlo evaluation of loop spectral
densities, was presented by Belova et. al. in ref. \cite{Yuri}.
  It was Ambj{\o}rn,
Olesen, and Peterson, in ref. \cite{AOP}, who noticed that dimensional
reduction implies that
the ratio of string tensions between quarks in different group representations
should equal the ratio of the corresponding quadratic Casimirs, since this
can be shown to be true in two dimensions.  In particular, for $SU(2)$
lattice gauge theory at weak couplings, the prediction is
\beq
  {\chi_j[I,J] \over \chi_{\oh}[I,J]} = {4 \over 3} j(j+1)
\label{caspred}
\eeq
where $\chi_j[I,J]$ is the Creutz ratio for Wilson loops in the
$j=0,\oh,1,{3\over 2},...$ representations.  These authors tested the
above prediction numerically, in both $D=3$ and $D=4$ dimensions, and
found it to be accurate to within 10\%.
Their results have since been confirmed, for larger loops and with
better statistics, by a number of other studies in both three and
four dimensions \cite{Casimirs}.
Similar results have also been obtained in $SU(3)$ gauge theory
\cite{Manfried}.  Of course, this ``Casimir scaling'' of string tensions
cannot hold at arbitrarily large distances, since at some distance the
screening of heavy quark charges by gluons
will become energetically favorable, reducing the effective
charge.  Numerical simulations indicate, however, that there is a large
distance interval, between the onset of confinement and the onset of
charge screening, where Casimir scaling of string tensions holds
quite accurately.\footnote{In fact, it is not even clear that color
screening has been seen yet, in lattice Monte Carlo simulations of
$D=3$ Yang-Mills theory, inside the scaling region (c.f. Poulis and
Trottier in \cite{Casimirs}).}
It is reasonable to demand that any theory of quark
confinement, which purports to explain the behavior of gauge fields beyond
the confinement scale, should account for the observed Casimir scaling of
interquark forces in this interval.

   Does the hypothesis of abelian dominance allow for the existence
of Casimir scaling?  According to a simple heuristic argument, found
in ref. \cite{Smit}, the answer is probably no.  Instead, beginning
at the onset of confinement, one expects
\beq
     \begin{array}{ll}
      \chi_j = \chi_{1/2} & ~~~~j={1\over 2},{3\over 2},{5\over 2},... \\ \\
      \chi_j = 0          & ~~~~j=1,2,3,...
     \end{array}
\label{monpred}
\eeq
for an $SU(2)$ gauge theory.
We refer to the expectations of eq. \rf{monpred} as the
``abelian monopole
prediction''.  We then test this prediction numerically in
two cases where one may be fairly sure that abelian monopole configurations
give the crucial contributions:

\begin{description}

\item{1a.} the calculation of Creutz ratios in lattice $D=3$ Yang-Mills
theory, using ``abelian-projected'' lattice configurations obtained
in maximal abelian gauge; and

\item{2a.} the calculation of Creutz ratios in the lattice $D=3$
Georgi-Glashow model in the Higgs phase.

\end{description}

\noindent The results for these two cases are compared with

\begin{description}

\item{1b.} the actual Creutz ratios in lattice $D=3$ Yang-Mills, obtained from
the full, unprojected lattice configurations; and

\item{2b.} Creutz ratios in the lattice $D=3$ Georgi-Glashow model
in the symmetric phase.

\end{description}

   It will be found that cases $1a$ and $2a$ agree quite well with the
abelian monopole prediction, and utterly disagree with the corresponding
cases $1b$ and $2b$, which instead follow the predictions of dimensional
reduction.  This has two consequences.  First, it means that
in the case of $GG_3$ in the Higgs phase, where it is known that an
effective $U(1)$ theory describes the infrared dynamics, the monopole
prediction is verified.  Secondly, in the case of pure Yang-Mills theory,
where Casimir scaling is observed, the abelian dominance approximation
has failed entirely.

\bigskip

   Before proceeding to discuss the simulations, let us first recall
the heuristic argument leading to the
abelian monopole prediction \rf{monpred}.  Suppose that, in an $SU(2)$
gauge theory, the area law for Wilson loops is due to fluctuations
of the gauge field $A^3_\m$,
associated with a remaining $U(1)$ symmetry.  This $U(1)$ symmetry is assumed
to be singled out either by an abelian-projection gauge choice (as in
't Hooft's theory), or by a unitary gauge choice ($D=3$
Georgi-Glashow model).  In that case, we would have
\bea
      <W_j(C)> &=& <\mbox{Tr}\exp [i \oint dx^\m A^a_\m T_a^j]>
\non \\
               &\sim& <\mbox{Tr}\exp [i \oint dx^\m A^3_\m T_3^j]>
\non \\
               &\sim& \sum_{m=-j}^j <\exp [im \oint dx^\m A^3_\m]>
\eea
where the $T^j_a$ are the $SU(2)$ group generators in the $j$-representation.
If an area law is obtained from abelian configurations, this is presumably
due to monopole effects.  Following Polyakov's analysis \cite{Poly}, one
then expects
\beq
      <W_j(C)> \sim \sum_{m=-j}^j \exp[-\m_m ~ \mbox{Area}(C)]
\label{Wj}
\eeq
The $\m_m$ will increase with the magnitude of the $U(1)$ charge,
which is given by $|m|$ (for $m=0$, $\m_0=0$).
The above sum would then be dominated by those terms which are falling most
slowly with increasing area, i.e. $m=\pm \oh$, for $j=$ half-integer,
and $m=0$, for $j=$ integer.  In this way, we arrive at the monopole
prediction \rf{monpred}.

   Now the behavior \rf{monpred} is, in fact, what one expects asymptotically,
due to charge screening.  The problem, however, is that according to the
argument above this behavior actually begins right at the confinement scale,
and has nothing whatever to do with the physics of charge
screening.\footnote{We have emphasized this lack of connection to charge
screening in the abelian projection theory in a
previous publication, which was mainly concerned with large-$N$ behavior
\cite{Us}. In the present article, we turn our attention to $N=2$.}
The fact that adjoint loops
are unconfined, in the abelian projection theory, is simply due to the
fact that the $m=0$ component of an adjoint charge is neutral (and thereby
unconfined) with respect to the remaining $U(1)$ symmetry. The $m=0$
contribution therefore dominates the sum in \rf{Wj}.\footnote{Of course,
if one would simply toss out the $m=0$ contribution, then $W_j(C)$
{\it would} decay exponentially with the area.  But we can see no
justification for such a procedure.}  A flux tube
between adjoint quarks doesn't form and then break due to charge screening;
in this picture the tube doesn't form at all.
As already mentioned, this conclusion appears to be contradicted by
the numerical evidence presented in
refs. \cite{AOP}, \cite{Casimirs}, and \cite{Manfried}, which find a
force between adjoint quarks which is about $8/3$ that of the
fundamental quarks, over a fairly large distance interval in the confinement
regime.

   The abelian monopole prediction, however, is based on a heuristic argument;
it could be that there is some subtlety of monopole dynamics that we
have missed.  Let us turn, then, to the numerical simulations.

\section{Breakdown of Abelian Dominance}

   We perform Monte Carlo simulations of $D=3$ lattice $SU(2)$ gauge theory,
at lattice coupling $\beta=5$, which is just inside the scaling regime.
Maximal abelian gauge-fixing, which maximizes the quantity
\beq
      Q_{sum} = \sum_{x,\m} \mbox{Tr}[U_\m(x) \s^3 U^\dagger_\m (x) \s^3]
\eeq
is implemented.  Wilson loops in the fundamental ($j=1/2$), adjoint
($j=1$), and $j=3/2$ representations, normalized to a maximum value of one,
are given by
\bea
       W_{\oh}(C) &=& \oh \mbox{Tr}[UUU....U]
\non \\
       W_{1}(C)   &=& {1 \over 3}(4W_{\oh}^2(C) - 1)
\non \\
       W_{\ot}(C) &=& {1\over 4}(8W_{\oh}^3(C) - 4W_{\oh}(C))
\label{reps}
\eea
We calculate the expectation values of these loops using both the
full link configurations (for which the gauge-fixing is irrelevant),
and also using the abelian-projected link configurations (or ``abelian
links'').  For a full $SU(2)$ link matrix, represented by
\beq
         U = a_0 I + i\sum_{k=1}^3 a_k \s^k
\eeq
the corresponding abelian link $U'$ is given by a truncation to the diagonal
component, followed by a rescaling to restore unitarity, i.e.
\beq
         U \ra U' = {a_0 I + i a_3 \s^3 \over \sqrt{a_0^2 + a_3^2} }
\label{abpro}
\eeq
Wilson loops of the abelian-projected configurations are obtained
by inserting the abelian links \rf{abpro} into eq. \rf{reps}, and the
corresponding Creutz ratios are computed in the usual way.

    Our simulation involved $100,000$ sweeps of a $12^3$ lattice
at $\b=5$, comprising $10,000$ thermalization sweeps, with data taken
every tenth of the remaining sweeps.  Figure 1 shows the ratios
of Creutz ratios
\beq
        {\chi_1[I,I] \over \chi_\oh[I,I]} ~~~~\mbox{and}~~~~
      {\chi_\ot[I,I] \over \chi_\oh[I,I]}
\eeq
for $I=2,3,4$.  The agreement with Casimir scaling (8/3 and 5,
respectively) is fairly good, as found in previous studies
\cite{AOP,Casimirs}.  Figure 2 shows the same ratio
of Creutz ratios, for the same loop sizes,
but this time computed with abelian-projected
configurations.  It is clear that Figures 1 and 2 display
completely different behavior.  In the abelian projection,
the adjoint Creutz ratio actually
goes negative at $I=3$; the adjoint tension is consistent with zero at $I=4$,
as predicted by \rf{monpred}.  Likewise, $\chi_\ot[I,I]$ appears
to converge to $\chi_\oh[I,I]$, again as expected from the
abelian monopole prediction.  However, this behavior of the abelian-projected
loops is clearly inconsistent with the corresponding behavior of the
full Wilson loops.  Evidently, for higher-representation Wilson loops,
abelian dominance has failed entirely.

\section{$D=3$ Lattice Georgi-Glashow Model}

   Polyakov's seminal work \cite{Poly} was concerned with the Higgs
phase of the Georgi-Glashow model in $D=3$ dimensions.  Because of this
work, we may be confident that the confinement mechanism in the Higgs phase is
due to monopole condensation.  In that case one may ask: is the monopole
prediction \rf{monpred} for higher representations confirmed?  And
does this prediction also hold in the symmetric phase?

    There have been a number of
lattice Monte Carlo simulations of this model, both in $D=3$ \cite{Nad,GG3}
and $D=4$ \cite{KLSW,GG4}
dimensions, and these have been mainly concerned with finding the
phase diagram of the theory.  To our knowledge, there has been no study
of the behavior of Wilson loops, as one goes across the symmetry-breaking
transition.  We have therefore carried out such a calculation.
However, as there is a three-dimensional coupling constant space
for the lattice Georgi-Glashow model, we have not attempted to compute
the Wilson loop behavior throughout the phase diagram.  Instead,
we have only computed loops along a particular line of the coupling
constant space, which crosses from the symmetric to the Higgs phase.
We believe the behavior that we find for Creutz ratios is typical, as
the system goes across the Higgs transition, but of course this will
have to be verified by a more extensive study.

   The lattice action of the Georgi-Glashow model is
\bea
     S &=& \oh \b_G \sum_{plaq} \mbox{Tr}[UUU^\dg U^\dg]
\non \\
       &+& \oh \b_H \sum_{n,\m} \mbox{Tr}[U_\m(n)\phi(n) U^{\dg}_\m(n)
                                                 \phi^\dg(n+\m)]
\non \\
       &-& \sum_n \left\{ \oh \mbox{Tr}[\phi \phi^\dg]
           + \b_R \left(\oh \mbox{Tr}[\phi \phi^\dg]-1 \right)^2 \right\}
\label{GGmodel}
\eea
where the adjoint Higgs field $\phi(n)$ has three degrees of freedom
per lattice site
\beq
        \phi(n) = i \sum_{a=1}^3 \phi^a(n) \s_a
\eeq
In performing the Monte Carlo simulations it is useful to go to a
unitary gauge where $\phi(n) = i \rho(n) \s_3$, reducing the degrees of
freedom of the Higgs field from three to one per site. The details
may be found in ref. \cite{Nad}.

   To map out the phase structure of the theory in $D=3$ dimensions,
we compute the following observables:
\begin{description}
\item{~~1.}  the rms value of the Higgs field
\beq
         R = <\mbox{Tr}[\phi \phi^\dg]>^{1/2}
\eeq

\item{~~2.} the value
\beq
        Q = \oh <\mbox{Tr}[U_\m(n) \s^3 U^\dg_\m(n) \s^3]>
\eeq
in unitary gauge.
\end{description}

\noindent A jump in these two quantities is an indication of a transition
from the symmetric phase to the Higgs phase.

   We begin by looking for a region of couplings where it is possible
to see a (fundamental) string tension in both the symmetric and
Higgs phases.  The strategy we have chosen is to keep $\beta_G$ and
$\beta_R$ fixed, and vary $\beta_H$.  One would like to use a value
of $\beta_G$ where the pure gauge theory is in the scaling regime,
i.e. $\beta_G \ge 5$.  In practice, however, we have not been able
to detect a string tension in the Higgs phase at such large values
of $\b_G$. Since presumably there {\it must} be a string tension in
the Higgs phase in $D=3$ dimensions, we interpret this result as meaning
that the monopole is quite heavy, in lattice units, at the larger values
of $\b_G$, and therefore the confinement scale (in the Higgs phase)
probably lies beyond the size of our
lattice.\footnote{A related observation has been made by Laursen and
M\"uller-Preussker in ref. \cite{GG3}, who noted that monopoles in the
Higgs phase, at $\b_G=5$, are very dilute.} So we have been forced to go to a
rather small value of $\b_G$, using $\b_G=2$ throughout.  A fixed
(and rather arbitrary)
value of $\beta_R=.01$ was also chosen; this was mainly in order to compare
our values for the location of the phase transition with those in ref.
\cite{Nad}.  Simulations in the region of the transition were run on
a $12^3$ lattice with a total of 35000 sweeps; of which 5000 were
thermalizing sweeps, with data taken every tenth of the remaining
sweeps.

   Fig. 3 shows the variation of the $Q$-parameter with $\beta_H$,
at fixed $\b_G=2,~\b_R=0.01$.  There is clear evidence of a 1st-order
transition between the symmetric and Higgs phases at $.45<\b_H<.46$;
which is supported by the behavior of the rms value of the Higgs field,
shown in Fig. 4, showing a similar jump at the same value of $\b_H$.

   Having located the transition to the Higgs phase, we then study
the behavior of Creutz ratios.  Fig. 5 shows the $\chi_j[2,2]$ Creutz
ratios for fundamental and adjoint loops.  Up to the Higgs transition,
we are in the strong-coupling regime and the Creutz
ratios do not appear to be strongly dependent on $\b_H$ (for
comparison, to lowest-order in the strong-coupling expansion at $\b_H=0$,
we have string tensions $\m_{1/2}=.84$ and $\m_1=2.01$).\footnote{The
lowest-order strong-coupling
result in three-dimensions is the same as that in two dimensions,
consistent with the idea of dimensional reduction, and the string
tension is given by a ratio of Bessel functions.  This ratio becomes,
in the limit of weak couplings, a ratio of quadratic Casimirs, which is
the origin of the Casimir scaling prediction of ref.
\cite{Poul}.}${}^,$\footnote{Creutz ratios for the $j=3/2$ representation
are not shown, since
the statistical errors are quite large for $2\times 2$ loops in the
symmetric phase.  However, we have found that the smaller
$1\times 1$ and $1\times 2$ loops are quite close to their
strong-coupling values in the symmetric phase, right up to the
transition.}  At the Higgs transition the
fundamental string tension drops, but remains finite, while the
adjoint string tension appears to be consistent with zero.

   In Fig. 6 we display the $\chi_j[I,I]$ Creutz ratios
in the Higgs phase, for the fundamental ($j=\oh$), adjoint ($j=1$),
and $j=3/2$ representations, at $I=2,3,4$.  The coupling is
$\b_H=.46$, which is just past the transition (once again,
$\b_G=2$ and $\b_R=.01$).  Note that the adjoint ratio
actually goes {\it negative} at $I=3$, and is
consistent  with
zero at $I=4$.  Since the signal for the $j=3/2$ loops is quite
small, we have not obtained good data for the $j=3/2$ Creutz
ratio beyond $I=3$.  Nevertheless, from the data at $I=2$ and
$I=3$, it does appear that the $j=3/2$ string tension is converging
to the $j=1/2$ value.

   In short, up to the Higgs transition, our Creutz ratios essentially
follow the strong-coupling expansion, which (at lowest order)
is in agreement with the notion of dimensional reduction.
At the Higgs transition,
both the absolute and relative values of the string tensions change
abruptly, and all indications are that the abelian monopole prediction
\rf{monpred} is fulfilled.

\section{Conclusions}

    At a minimum, our results cast considerable doubt on the
hypothesis of abelian dominance in maximal abelian gauge.  If the
``photon'' gauge field associated with the remaining $U(1)$ symmetry
is mainly responsible for forces between heavy fundamental quarks beyond
the confinement scale, that same gauge field should also explain the forces
between heavy quarks in higher group representations.  Given that
the projection to abelian lattice configurations is found to reproduce
the fundamental string tension, then according to these ideas
the string tensions for higher representations
should also be reproduced, at any distance beyond the confinement
scale. We have found, however, that this is not at all the case.

    There have been previous indications of trouble for the
abelian projection theory.  As three of us have pointed out in a previous
publication \cite{Us}, for $SU(N)$ theories
there is a significant difference in the coefficients of
subleading (perimeter-law) contributions to adjoint  Wilson
loops, as predicted, respectively, by large-$N$ counting
arguments, and by the abelian-projection
theory.  The origin of this difference is that according to the
large-$N$ picture, the perimeter-law term is due to the binding
of gluons to the adjoint quarks (a $1/N^2$ suppressed process),
while perimeter law behavior in the abelian projection theory is just
due to the fact that $N-1$ of the $N^2-1$ adjoint quark charges are
neutral with respect to the abelian subgroup, and this leads only
to a $1/N$ suppression factor.  The different powers of $N$
reflect the fact that there are different mechanisms involved; only
one of these can be the right explanation of the perimeter law.
We refer the reader to ref. \cite{Us} for a more extensive discussion
of this point.  Some other types of numerical evidence against the
abelian projection theory are found in ref. \cite{Junichi}.

   Not everyone finds large-$N$ arguments persuasive,  so in this
article we have considered the opposite limit, namely $N=2$.
For such a small value of $N$, it is hard to understand, in the context of
the abelian projection theory, why the abelian neutral ($m=0$) adjoint quark
component should not completely dominate the value of the adjoint loop, at
and beyond the confinement scale.  In fact we find, in abelian projected
lattice configurations, that this is exactly what happens, and the
corresponding adjoint loop has no discernable string tension at any of the
distances studied.  However, such behavior is in complete contrast to adjoint
Creutz ratios, measured at the same distances, constructed from the
full lattice configuration.  The latter follow Casimir scaling \rf{caspred}.
The breakdown of abelian dominance in pure $SU(2)$ lattice gauge theory,
not only for the adjoint but also for the $j=3/2$ representations, seems to
be quite evident from comparing Figs. 1 and 2.
Conversely, in the $D=3$ Georgi-Glashow model in the Higgs phase, where the
infrared dynamics is essentially that of compact QED, it is the monopole
prediction, rather than Casimir scaling, which agrees with the data.

   A breakdown of abelian dominance implies that large scale vacuum
fluctuations are {\it not} adequately represented by the fluctuations of only
those degrees of freedom associated with a particular Cartan subalgebra
($A^3_\m$, in the Yang-Mills case considered
here), not even in the maximal abelian gauge.  Large-scale fluctuations in the
``off-diagonal'' degrees of freedom ($A^1_\m$ and $A^2_\m$) have been found
to be important; were it not for these fluctuations, Wilson loops would
follow the abelian monopole prediction found for abelian projected
configurations.  It may be, of course, that there exists a simple effective
theory, perhaps even an abelian gauge theory involving some sort of
composite fields, which does capture the
essential dynamics of confinement in Yang-Mills theory.
It may also be that the Yang-Mills vacuum does, in some way, exhibit
the properties of a dual-superconductor.  Concerning
these possibilities, we have nothing to say here.  What can be asserted,
however,
is that an effective theory of the long-wavelength dynamics cannot be
based on the $A^3_\m$ degrees of freedom alone.  The validity of a
theory of that sort would imply the validity of the abelian dominance
approximation, and this simply conflicts with our data.

   Some caveats about the data, however, are in order.  We have looked
only at rather small loops (up to $4\times 4$ lattice spacings) at $\b=5$
in $D=3$ pure Yang-Mills, and only along a single line (varying one
coupling) in the 3-dimensional phase
diagram of the $D=3$ Georgi-Glashow model.  Certainly much more numerical
work is needed to extend and solidify our results.  This work is in
progress, and will be reported in due course.

    Finally, in view of the observed Casimir scaling of Creutz ratios,
we believe that a certain scepticism regarding proposed monopole
confinement mechanisms, at least in their most naive forms, may be appropriate.
Whatever may be the importance of monopoles, it appears doubtful that
the effective infrared dynamics of Yang-Mills theory is essentially that of
compact QED.  It may also be that there is an element of truth
in some of the old ideas regarding dimensional reduction.
In any event, Casimir scaling of heavy interquark
forces is a striking result of many numerical simulations, and any
satisfactory theory of quark confinement must eventually take this scaling
into account.

\bigskip
\bigskip

\noindent {\sl Note Added : } After submitting the present paper
for publication, a paper appeared by Poulis \cite{P} which also addresses
the problem of abelian dominance for higher representation sources.
Poulis modifies the usual abelian dominance approximation,
in an attempt to allow for some of the effects of the off-diagonal 
degrees of freedom,
and finds that in this modified approximation the $m=0$ adjoint loop
component still has no area law falloff in any distance range.
In our opinion his results support our conclusions, 
although he chooses to interpret those results in a 
different way.  We will return to this issue in a future publication
\cite{Us2}.

\vspace{33pt}

\noindent {\Large \bf Acknowledgements}

\bigskip

  J.G. is grateful for the hospitality of the Niels Bohr
Institute, where some of this work was carried out.  He would also like to
thank Thomas Kaeding for his kind assistance in preparing the postscript
figure files.

  This research was supported in part by Fonds zur F\"orderung
der wissenschaftlichen Forschung under Contract P7237-TEC (M.F.), the
U.S. Dept. of Energy, under Grant No. DE-FG03-92ER40711 (J.G.), and
Bundes\-minis\-terium f\"ur Wissen\-schaft und For\-schung and Slovak
Grant Agency for Science, Grant No. 2/1157/94 ({\v S}.O.).

\newpage

\noindent {\Large \bf Figure Captions}

\bigskip
\bigskip

\begin{description}

\item[Fig. 1] The ratio of Creutz ratios $\chi_j[I,I]/\chi_\oh[I,I]$,
for $j=1$ (solid circles) and $j=3/2$ (open circles), in $D=3$ lattice
$SU(2)$ gauge theory, at $\b=5$.  Dashed lines show the corresponding ratio
of quadratic Casimirs (${8 \over 3}$ for $j=1$, and $5$ for $j=3/2$).

\item[Fig. 2] Same as Fig. 1, except that Creutz ratios have been
computed using the abelian-projected lattice configurations in maximal
abelian gauge.

\item[Fig. 3]  Variation of the $Q$-parameter with $\b_H$ in the
3D Georgi-Glashow model, at $\b_G=2$ and $\b_R=.01$,

\item[Fig. 4] Variation of the rms Higgs field $R$ with $\b_H$,
same model and parameters as in Fig. 3.

\item[Fig. 5] Creutz ratios $\chi_j[2,2]$ vs. $\b_H$,
for $j=\oh$ (solid circles) and $j=1$ (grey squares), same model and
parameters as in Fig. 3.

\item[Fig. 6] Creutz ratios $\chi_j[I,I]$ vs. $I$, in the
Higgs phase of the 3D Georgi-Glashow model at $\b=.46$, just
past the transition. Again $\b_G=2$, $\b_R=.01$; representations $j=\oh$
(solid circles), $j=1$ (grey squares), and $j=3/2$ (open circles) are shown.

\end{description}


\newpage

\begin{thebibliography}{xx}
\bibitem{Poly} A. Polyakov, Nucl. Phys. B120 (1977) 429.
\bibitem{thooft} G. 't Hooft, Nucl. Phys. B190 [FS3] (1981) 455.
\bibitem{th2} G. 't Hooft, {\it in}: High Energy Physics, ed.
A. Zichichi (Editrice Compositori, Bologna, 1976).
\bibitem{Mand} S. Mandelstam, Phys. Reports 23C (1976) 245.
\bibitem{Suzuki} T. Suzuki and I. Yotsuyanagi, Phys. Rev. D42 (1990) 4257; \\
 S. Hioki et al., Phys. Lett. B272 (1991) 326.
\bibitem{Luigi} L. Del Debbio, A. Di Giacomo, G. Paffuti, and
P. Pieri, Phys. Lett. B355 (1995) 255.
\bibitem{KLSW} A. Kronfeld, M. Laursen, G. Schierholz, and
U.-J. Wiese, Phys. Lett. B198 (1987) 516.
\bibitem{JG} J. Greensite, Nucl. Phys. B166 (1980) 113; Nucl. Phys. B158
(1979) 469.
\bibitem{Poul} P. Olesen, Nucl. Phys. B200 [FS4] (1982) 381.
\bibitem{Yuri} T. Belova, Yu. Makeenko, M. Polikarpov, and A. Veselov,
Nucl. Phys. B230 [FS10] (1984) 473.
\bibitem{AOP} J. Ambj{\o}rn, P. Olesen, and C. Peterson, Nucl. Phys. B240
[FS12] (1984) 189; 533.
\bibitem{Casimirs} G. Poulis and H. Trottier, archive: hep-lat/9504015; \\
C. Michael, Nucl. Phys. B (Proc. Suppl.) 26 (1992) 417;
Nucl. Phys. B259 (1985) 58.
\bibitem{Manfried}
N. Cambell, I. Jorysz, and C. Michael, Phys. Lett. B167 (1986) 91; \\
M. Faber and H. Markum, Nucl. Phys. B (Proc. Suppl.) 4 (1988) 204; \\
M. M\"uller, W. Beirl, M. Faber, and H. Markum,  Nucl. Phys. B
(Proc. Suppl.) 26 (1992) 423.
\bibitem{Smit} J. Smit and A. van der Sijs, Nucl. Phys. B355 (1991) 603.
\bibitem{Us} L. Del Debbio, M. Faber, and J. Greensite, Nucl. Phys. B414
(1994) 594.
\bibitem{Nad} S. Nadkarni, Nucl. Phys. B334 (1990) 559.
\bibitem{GG3} M. Laursen and M. M\"uller-Preussker, Nucl. Phys.
B313 (1989) 1; \\
K. Farakos and G. Koutsoumbas, Z. Phys. C43 (1989) 301.
\bibitem{GG4} V. Bornyakov et al., Z. Phys. C42 (1989) 633; \\
G. Schierholz, J. Seixas, and M. Teper, Phys. Lett. B157 (1985) 209;
Phys. Lett. B151 (1985) 69.
\bibitem{Junichi} J. Greensite and J. Iwasaki, Phys. Lett. B255 (1991) 415.
\bibitem{P} G. Poulis, hep-lat/9601013.
\bibitem{Us2} L. Del Debbio, M. Faber, J. Greensite, and 
{\v S}. Olejn\'{\i}k, in preparation.
\end{thebibliography}
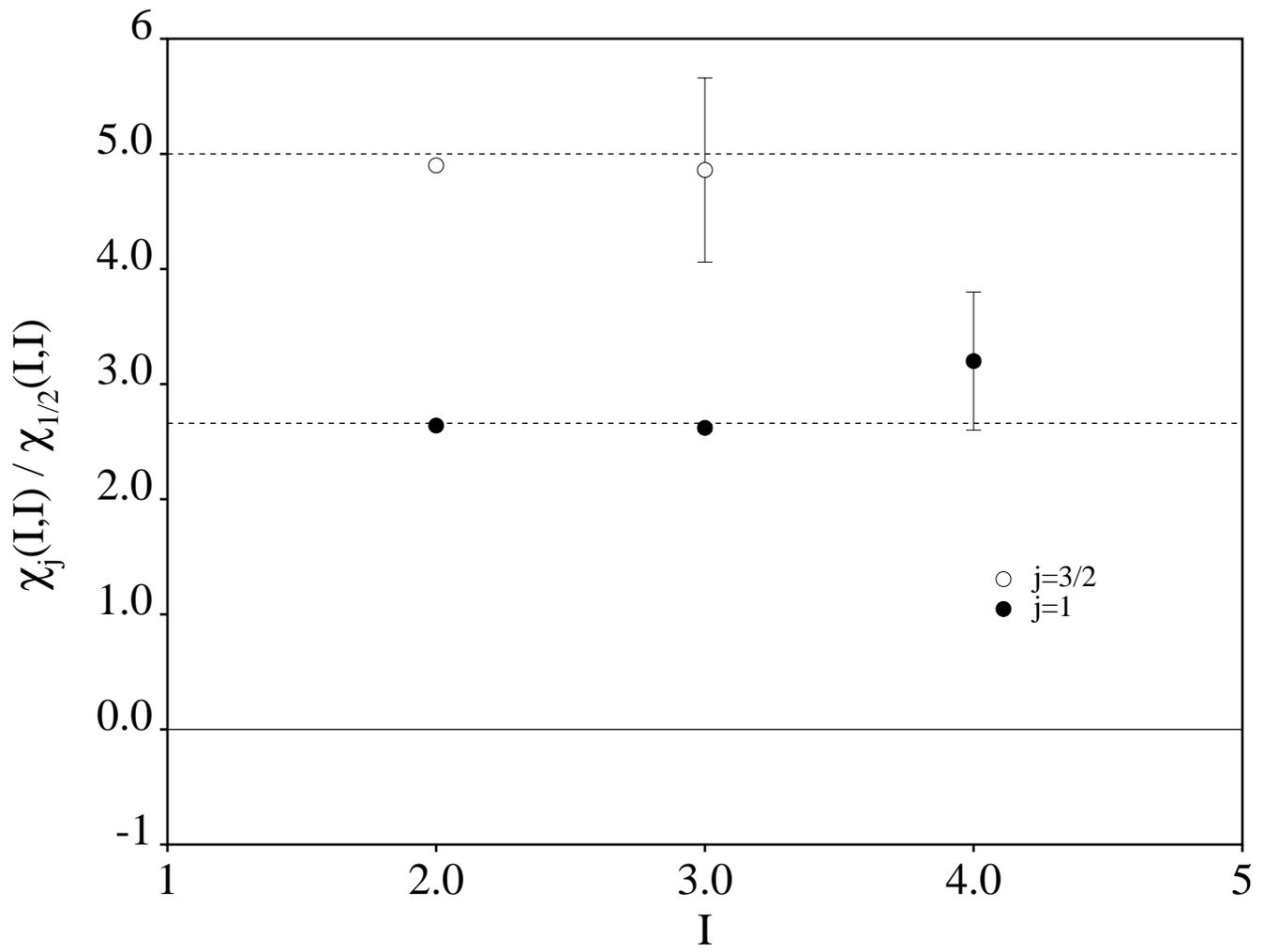
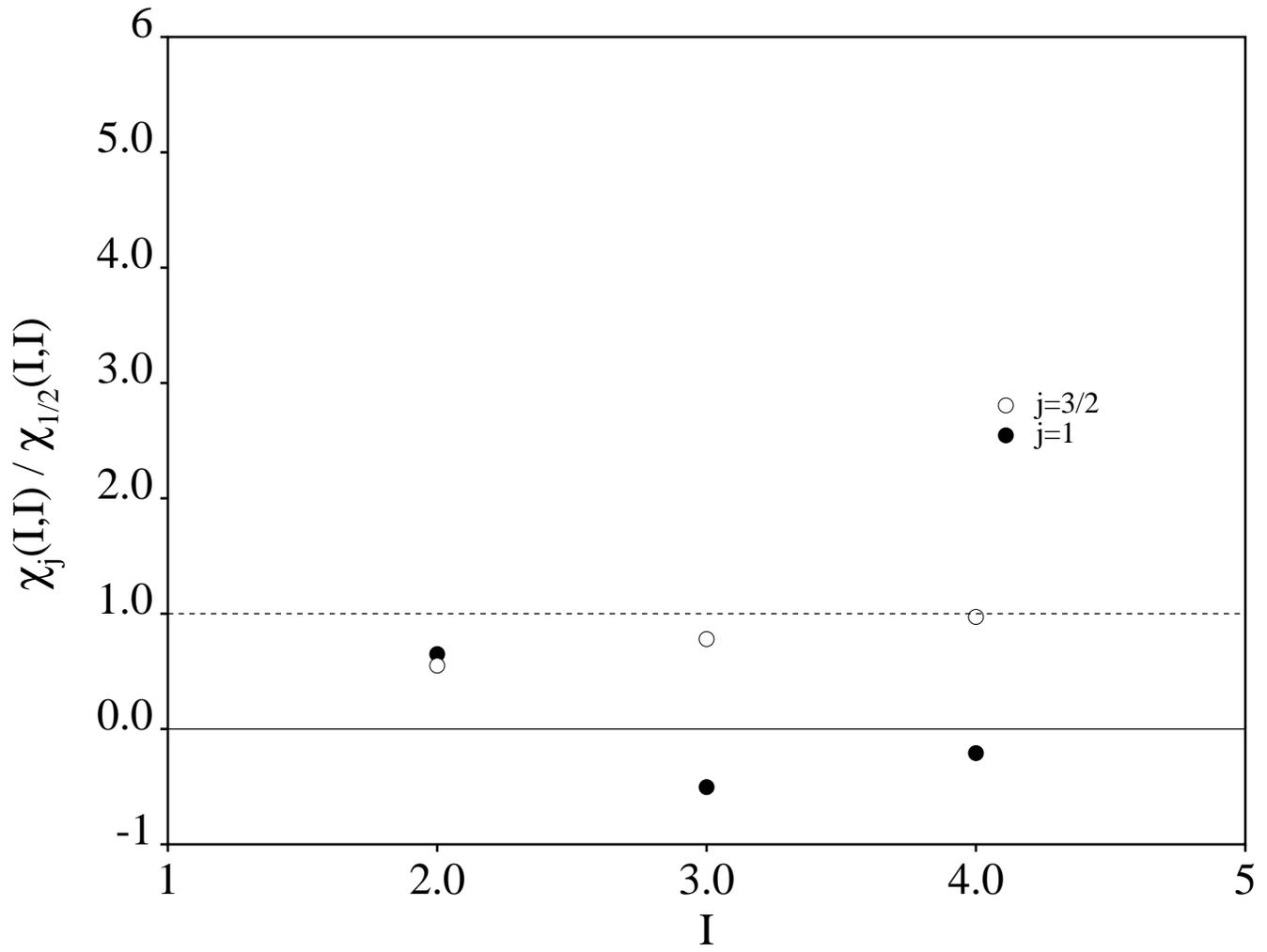
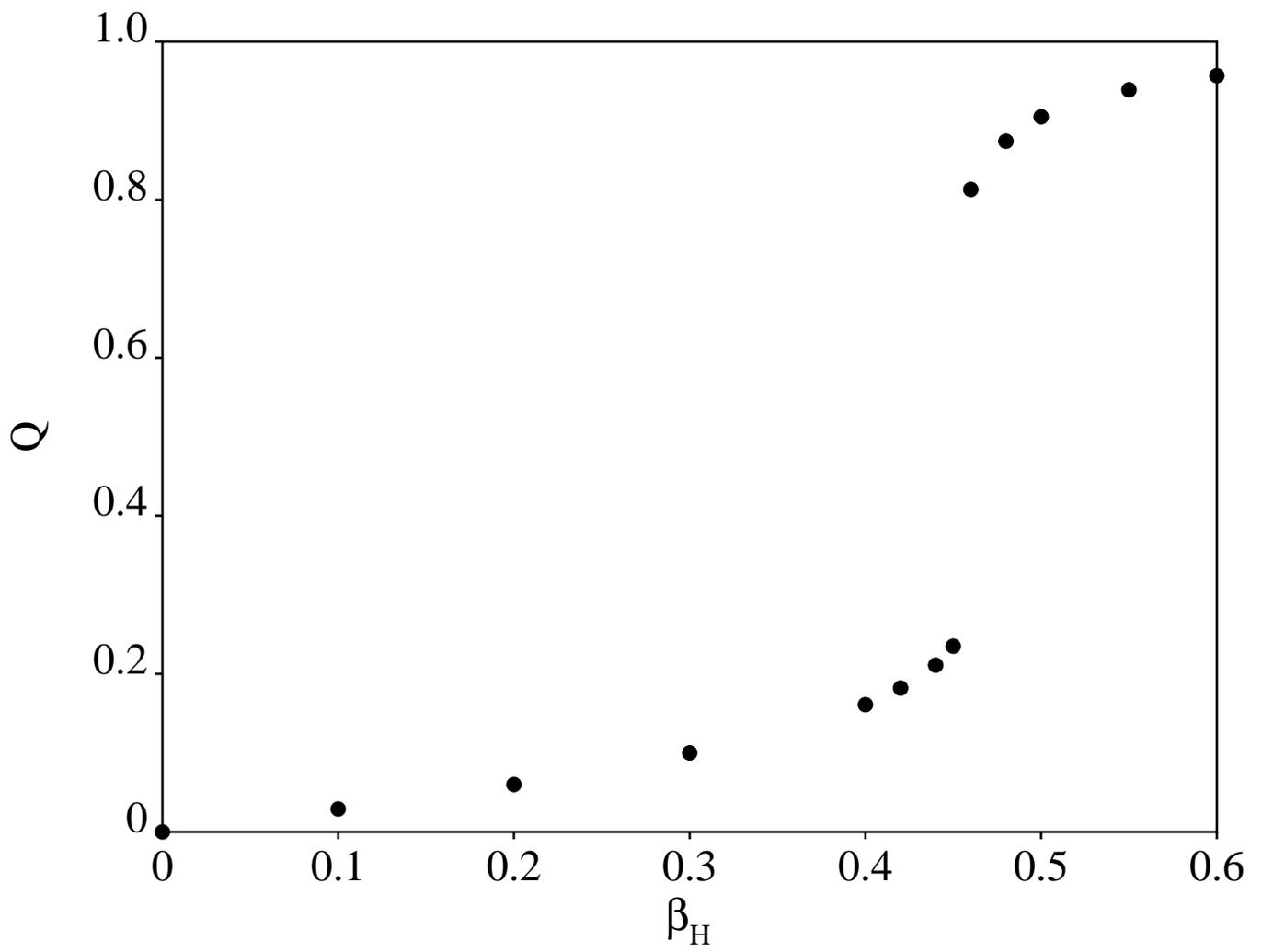
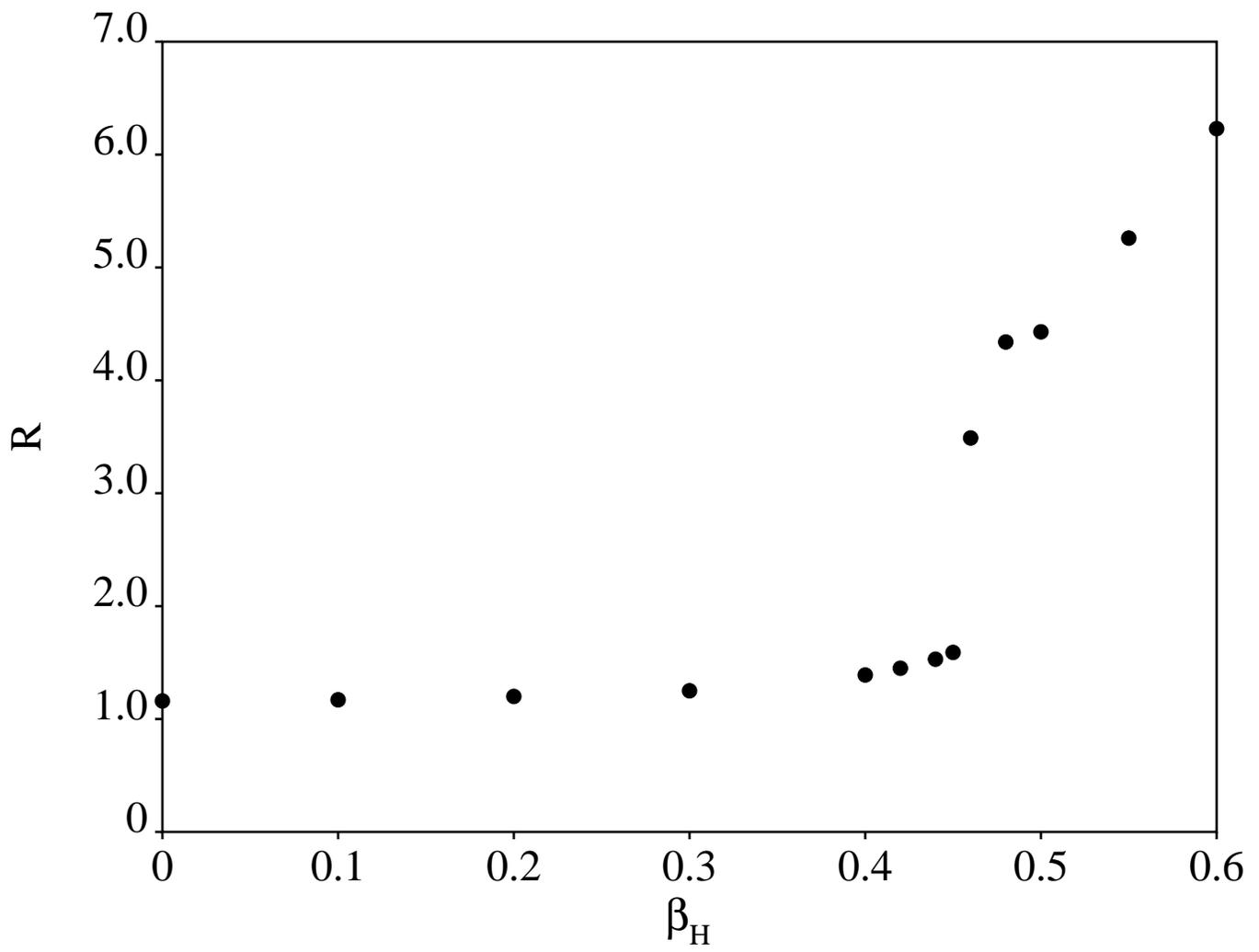
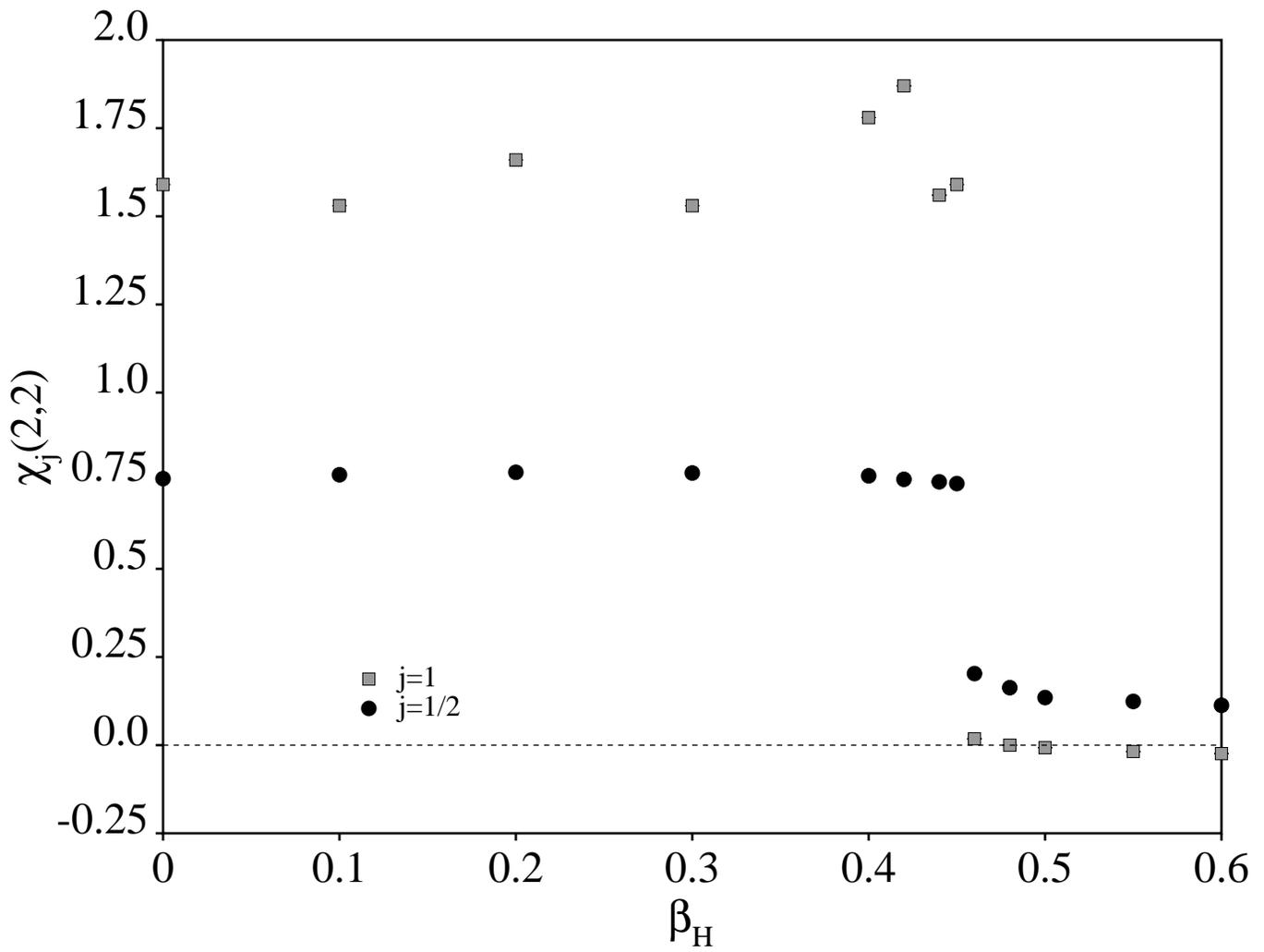
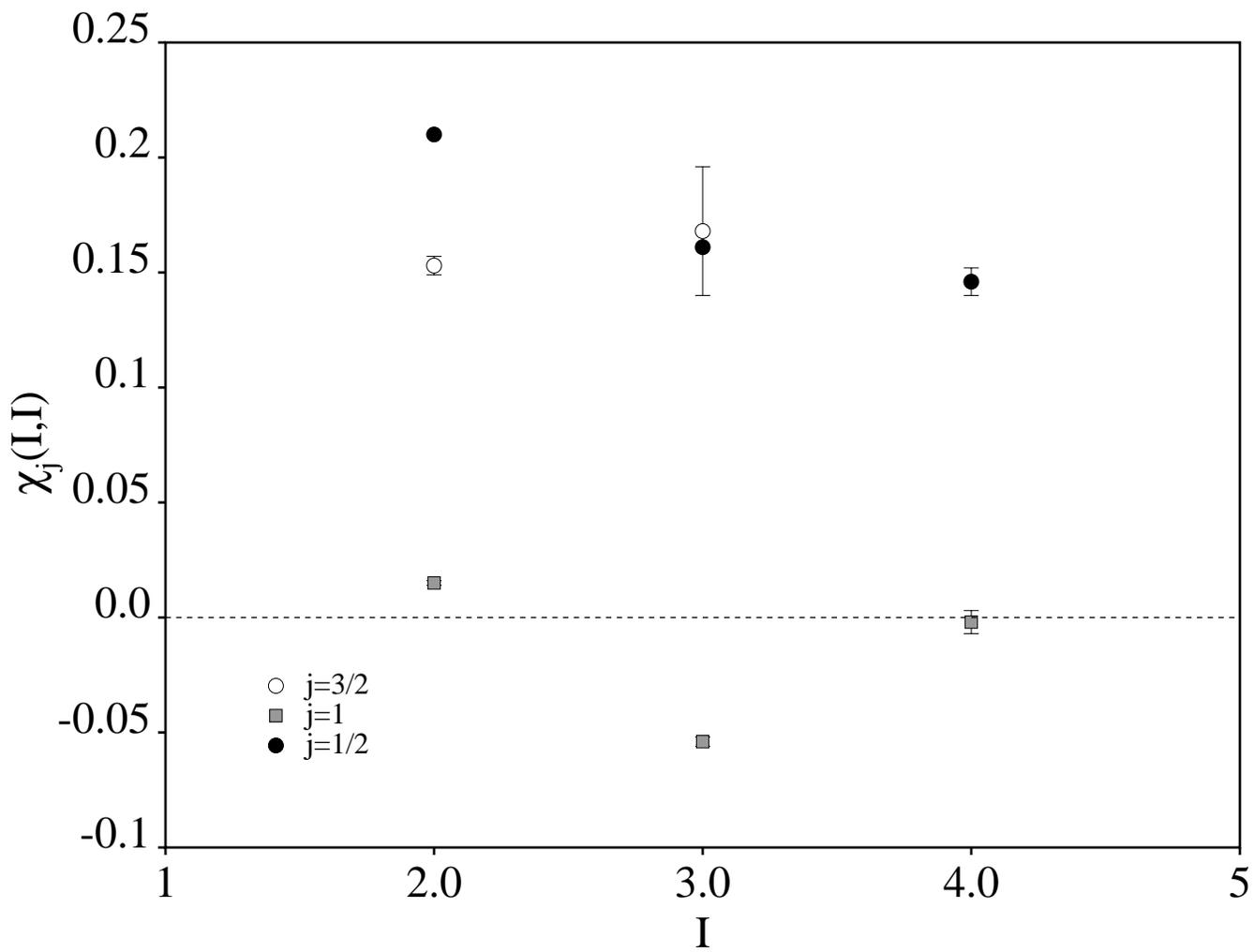
\end{document}